# Robust Multitask Diffusion Normalized M-estimate Subband Adaptive Filtering Algorithm Over Adaptive Networks

Wenjing Xu[1], Haiquan Zhao[1,*], Shaohui Lv[1]

**Abstract:** In recent years, the multitask diffusion least mean square (MD-LMS) algorithm has been extensively applied in the distributed parameter estimation and target tracking of multitask network. However, its performance is mainly limited by two aspects, i.e, the correlated input signal and impulsive noise interference. To overcome these two limitations simultaneously, this paper firstly introduces the subband adaptive filter (SAF) into the multitask network. Then, a robust multitask diffusion normalized M-estimate subband adaptive filtering (MD-NMSAF) algorithm is proposed by solving the modified Huber function based global network optimization problem in a distributed manner, which endows the multitask network strong decorrelation ability for correlated inputs and robustness to impulsive noise interference, and accelerates the convergence of the algorithm significantly. Compared with the robust multitask diffusion affine projection M-estimate (MD-APM) algorithm, the computational complexity of the proposed MD-NMSAF is greatly reduced. In addition, the stability condition, the analytical expressions of the theoretical transient and steady-state network mean square deviation (MSD) of the MD-NMSAF are also provided and verified through computer simulations. Simulation results under different input signals and impulsive noise environment fully demonstrate the performance advantages of the MD-NMSAF algorithm over some other competitors in terms of steady-state accuracy and tracking speed.

## 1. Introduction

Distributed adaptive estimation is an important technique for signal processing within a network, which performs specific tasks through continuous learning and adaptation of interconnected nodes in network. It has attracted extensive research due to its reliability, scalability, and resource efficiency [1]-[3]. Previous research on distributed estimation mainly focused on the distributed single-task network, the whole network only needs to estimate a single parameter vector. In this work, the distributed estimation of the cluster multitask network is considered, which is different from the single-task network in that there are multiple unknown parameter vectors need to be estimated. All nodes in the multitask network are divided into different clusters, each cluster has its own task (parameter vector to be estimated) [4]-[7].

---

[1]Wenjing Xu, Haiquan Zhao and Shaohui Lv are with the Key Laboratory of Magnetic Suspension Technology and Maglev Vehicle, Ministry of Education, and the School of Electrical Engineering, Southwest Jiaotong University, Chengdu,610031,China.
*Corresponding author
E-mail:wenjingxv@126.com; hqzhao@home.swjtu.edu.cn; shaohuilv_swjtu@126.com;

In recent years, the multitask diffusion LMS (MD-LMS) algorithm has been well applied to the parameter estimation and target tracking of the multitask networks [6]. However, its performance is mainly limited by two aspects: On the one hand, its convergence speed largely depends on the characteristics of the input signal. Specifically, the large eigenvalue spread of the correlation matrix of the correlated input signal will lead to slow convergence of the algorithm. On the other hand, the optimality of the algorithm based on the MMSE criterion depends on the Gaussian noise assumption to a large extent. When the measurement noise does not conform to the Gaussian distribution, the performance of the MD-LMS algorithm will deteriorate significantly.

In view of the former defect, reference [8] proposes the multitask diffusion affine projection algorithm (MD-APA) to improve the convergence speed of the algorithm under the correlated inputs. In fact, the MD-APA is a generalization of the normalized version of MD-LMS in the time domain. For the latter defect, on the basis of MD-APA, the multitask diffusion affine projection M-estimate (MD-APM) algorithm is proposed by applying the M-estimate function to remove outliers [9], [10]-[13]. The multitask diffusion affine projection maximum correntropy criterion (MD-APMCC) algorithm is also proposed by introducing the maximum correntropy criterion (MCC) into the MD-APA [14], [15], which is insensitive to large outliers. These algorithms are all robust to impulsive noise interference.

Unfortunately, the above robust multitask algorithms are all based on the AP strategy, which not only improve the convergence speed of the algorithm, but also greatly increase the computational complexity. Inspired by the subband adaptive filter (SAF) [16]-[22], we extend the normalized version of the MD-LMS algorithm to the subband domain, and introduce the M-estimate function with the property of removing outliers, and propose a robust multitask diffusion normalized M-estimate subband adaptive filtering (MD-NMSAF) algorithm. The proposed MD-NMSAF algorithm inherits the strong decorrelation ability of SAF, improves the convergence speed of the algorithm under correlated inputs and endows the distributed network robustness to impulsive noise. Furthermore, compared with the MD-APA, the structure of SAF can greatly reduces the computational complexity of the MD-NMSAF algorithm. The main contributions of this paper are summarized as follows:

1) The cluster multitask diffusion algorithm is extended to the subband domain, and the M-estimate function is introduced to propose a robust multitask diffusion SAF algorithm, i.e, MD-NMSAF algorithm, which has the strong decorrelation ability for correlated inputs and robustness to impulsive noise interference.
2) The mean and mean square stability of the MD-NMSAF algorithm are analyzed in impulsive noise environment, and the stability conditions are also given. Then, the analytical expressions of the theoretical transient and steady-state network network mean square deviation (MSD) are provided.

3) The computational complexity of the proposed MD-NMSAF algorithm with some other multitask diffusion algorithms are summarized in the form of table, and the computational advantages of the MD-NMSAF algorithm over the AP based multitask diffusion algorithm are explained in detail.
4) The influences of the step size and subband number on the performance of MD-NMSAF are studied, and the validity and accuracy of the mean square stable step size upper bound, theoretical transient and steady-state analysis results are verified under different input signals.
5) The convergence and tracking performance between the proposed MD-NMSAF algorithm and some other multitask diffusion algorithms are compared in different input signals, the superiority of the MD-NMSAF algorithm is verified.

The remainder of this paper is organized as follows. In section 2, the MD-NMSAF algorithm is derived in detail. In section 3, the mean and mean-square stability, theoretical transient and steady-state performance and computational complexity of the proposed MD-NMSAF algorithm are analyzed. Computer simulations are provided in Section 4. Finally, the conclusion is given in Section 5.

## 2. The proposed algorithm

Consider a clustered multitask network consists of $N$ nodes, the nodes on this network are divided into $L$ clusters ($L$ tasks). Node $k$ has access to the reference signal $d_k(t)$ and input signal vector $\boldsymbol{u}_k(t)$, both of them are related by the following linear model:

$$d_k(t) = \boldsymbol{u}_k^T(t)\boldsymbol{w}_k^* + v_k(t) \tag{1}$$

where the $M \times 1$ vector $\boldsymbol{w}_k^*$ is the target parameter vector of node $k$, and $v_k(n)$ denotes the measurement noise at node $k$. The target parameter vectors of nodes in the same cluster $\mathcal{C}_l, l = 1,2,\ldots L$ are the same, i.e., $\boldsymbol{w}_k^* = \boldsymbol{w}_{\mathcal{C}_l}^*$, for $k \in \mathcal{C}_l$. In addition, it is assumed that there are some similarities between the target parameter vectors of neighboring clusters [4].

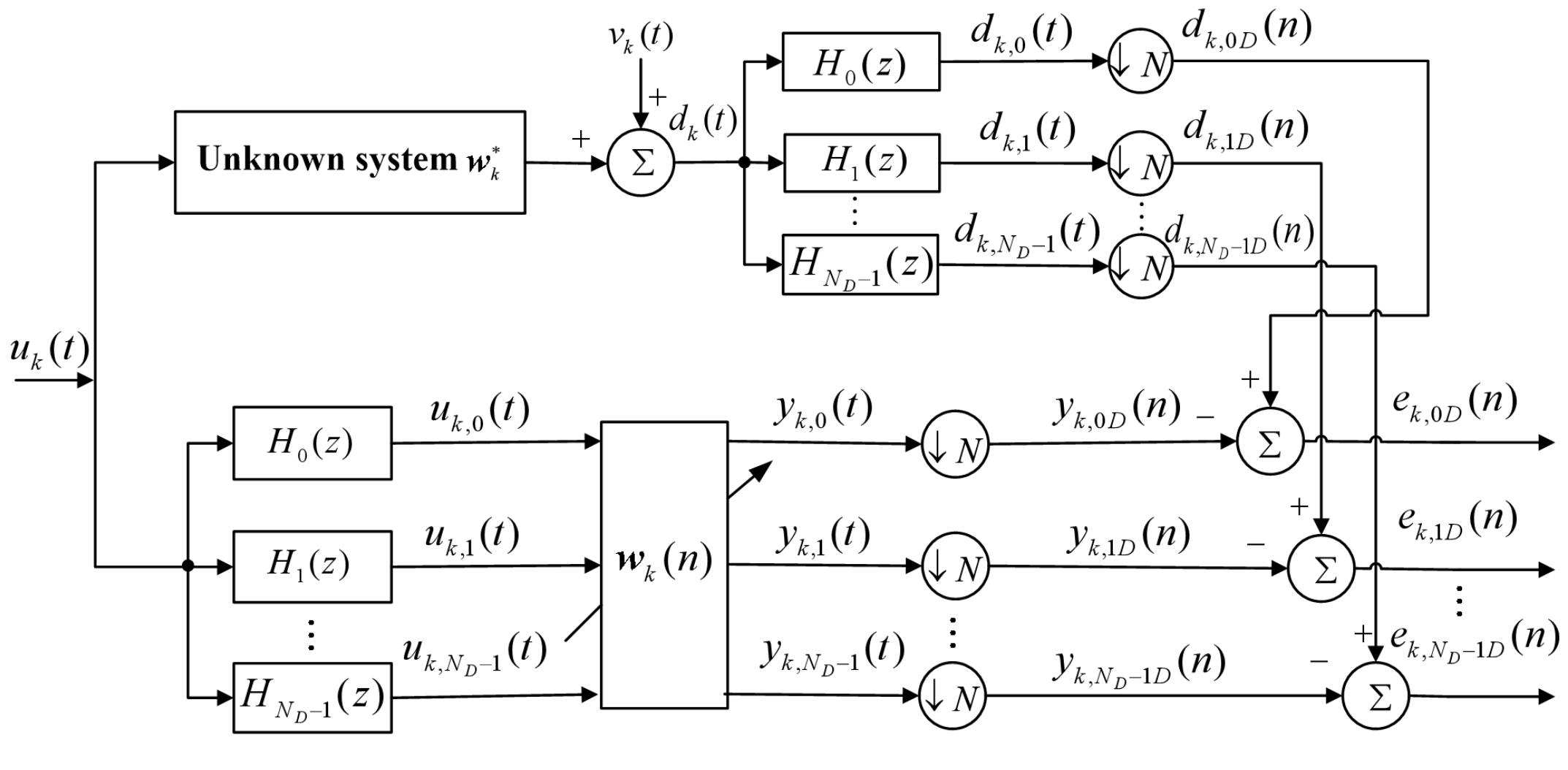

Fig.1. Multiband structure of subband adaptive filter.

In order to accelerate the convergence of the MD-LMS algorithm under highly correlated input, we introduce the SAF into the multitask network and reduces the correlation of input signal by subband dividing and extracting [16]. Fig.1 shows the multiband-structured SAF with $N_D$ subbands. For each node *k*, the reference signal $d_k(t)$ and input signal vector $\boldsymbol{u}_k(t)$ are partitioned into $N_D$ band-dependent signals $d_{k,i}(t)$ and $\boldsymbol{u}_{k,i}(t)$ by the analyzing filter bank $\{H_i(z), i = 0,1,\ldots,N_D - 1\}$, respectively. The subband output signal $y_{k,i}(t)$ is obtained by inputting the subband signal $\boldsymbol{u}_{k,i}(t)$ into the SAF with tap-weight vector $\boldsymbol{w}_k(n) = [w_{k,1}(n), w_{k,2}(n), \ldots, w_{k,M}(n)]^T$. Then, $d_{k,i}(t)$ and $y_{k,i}(t)$ are critically decimated to $d_{k,iD}(n) = d_{k,i}(nN_D)$ and $y_{k,iD}(n) = y_{k,i}(nN_D)$, respectively. The variables *t* and *n* represent the index of the original and decimated sequences, respectively. Therefore, the decimated subband error signal $e_{k,iD}(n)$ can be expressed as

$$\begin{aligned} e_{k,iD}(n) &= d_{k,iD}(n) - y_{k,iD}(n) \\ &= d_{k,iD}(n) - \boldsymbol{u}_{k,i}^T(n)\boldsymbol{w}_k(n) \end{aligned} \tag{2}$$

where $\boldsymbol{u}_{k,i}(n) = \left[\boldsymbol{u}_{k,i}(nN_D), \boldsymbol{u}_{k,i}(nN_D - 1), \ldots, \boldsymbol{u}_{k,i}(nN_D - M + 1)\right]^T$. Similar to the MD-LMS algorithm [6], to solve a multitask network global optimization problem in a distributed manner, the cost function must rely on the local interaction of measurement data within neighborhoods and the similarity relationship between adjacent tasks. At the same time, after introducing the M-estimate function, we can construct the following local cost function

$$\begin{aligned} J_{\mathcal{C}(k)}(\boldsymbol{w}_k) &\triangleq J_k^{\text{loc}}(\boldsymbol{w}_k) + \sum_{l \in \mathcal{N}_k^- \cap \mathcal{C}(k)} J_l^{\text{loc}}(\boldsymbol{w}_l) \\ &= \frac{1}{2} \sum_{r \in \mathcal{N}_k \cap \mathcal{C}(k)} c_{k,r} \sum_{i=0}^{N_D - 1} \frac{\phi\left[d_{r,iD}(n) - \boldsymbol{u}_{r,i}^T(n)\boldsymbol{w}_k\right]}{\left\|\boldsymbol{u}_{r,i}(n)\right\|^2} + \frac{\eta}{2} \sum_{l \in \mathcal{N}_k \setminus \mathcal{C}(k)} \gamma_{k,l} \left\|\boldsymbol{w}_k - \boldsymbol{w}_l\right\|_2^2 + \sum_{m \in \mathcal{N}_k^- \cap \mathcal{C}(k)} b_{m,k} \left\|\boldsymbol{w}_k - \boldsymbol{w}_m^0\right\|_2^2 \end{aligned} \tag{3}$$

where $\mathcal{N}_k$ represents the set of neighbor nodes of node *k* (including itself), $\mathcal{N}_k^-$ is the set that $\mathcal{N}_k$ excluding node *k*, the symbol $\cap$ denotes the set intersection, $\setminus$ is the set difference, and $\mathcal{C}(k)$ denotes the cluster where node *k* is located. $c_{k,r}$ and $\gamma_{k,l}$ are non-negative intra-task and inter-task combination coefficients, respectively. $b_{m,k}$ is a non-negative coefficient, please refer to [2] for details. $\eta$ is called the regularization strength [6]. $\boldsymbol{w}_m^0$ is the target parameter vector of node *m*. Note that the M-estimate function $\phi[\cdot]$ adopted in this paper is the modified Huber function [10], given by

$$\phi[e_{k,iD}(n)] = \begin{cases} e_{k,iD}^2(n)/2, & \left|e_{k,iD}(n)\right| < \xi_{k,iD}(n) \\ \xi_{k,iD}^2(n)/2, & \text{other} \end{cases} \tag{4}$$

where $\xi_{k,iD}(n)=k_{\xi}\sigma_{k,iD}(n)$ is a time-varying threshold. $k_{\xi}$ is used to control the suppression effect of impulsive noise, and the typical value 2.576 is chosen in this paper. $\sigma_{k,iD}(n)$ is the standard deviation of the estimation error without impulsive noise, and can be estimated by $\sigma^2_{k,iD}(n)=\gamma\sigma^2_{k,iD}(n-1)+C(1-\gamma)med(A_{e_{k,iD}}(n))$, $0\ll\gamma<1$ is the forgetting factor; $C=1.483(1+5/(N_w-1))$ is a finite sample correction factor; $med(\cdot)$ denotes the median operator; $A_{e_{k,iD}}(n)=\left\{e^2_{k,iD}(n),\ldots,e^2_{k,iD}(n-N_w+1)\right\}$ is the sliding data window of length $N_w$, the value of $N_w$ must be adjusted with the change of impulsive noise environment. In our simulations, we set $N_w=5$.

Taking the derivative of the local cost function (3) with respect to $\boldsymbol{w}_k$, we can get

$$\begin{aligned}\nabla J_{\mathcal{C}(k)}(\boldsymbol{w}_k)&=\frac{\partial J_{\mathcal{C}(k)}(\boldsymbol{w}_k)}{\partial\boldsymbol{w}_k}\\&=-\frac{1}{2}\sum_{r\in\mathcal{N}_k\cap\mathcal{C}(k)}c_{k,r}\sum_{i=0}^{N_D-1}\frac{\varphi\left[d_{r,iD}(n)-\boldsymbol{u}^T_{r,i}(n)\boldsymbol{w}_k\right]}{\left\|\boldsymbol{u}_{r,i}(n)\right\|^2}\boldsymbol{u}_{r,i}(n)\\&\quad+\eta\sum_{l\in\mathcal{N}_k\setminus\mathcal{C}(k)}\gamma_{k,l}\left(\boldsymbol{w}_k-\boldsymbol{w}_l\right)+2\sum_{m\in\mathcal{N}_k^-\cap\mathcal{C}(k)}b_{m,k}\left(\boldsymbol{w}_k-\boldsymbol{w}^0_m\right)\end{aligned}\tag{5}$$

where $\varphi[\cdot]=\partial\phi[\cdot]/\partial[\cdot]$. By applying the stochastic gradient descent method, at instant *n*, the weight vector of the MD-NMSAF algorithm at node *k* is updated as

$$\begin{aligned}\boldsymbol{w}_k(n+1)&=\boldsymbol{w}_k(n)-\mu\nabla J_{\mathcal{C}(k)}(\boldsymbol{w}_k(n))\\&=\boldsymbol{w}_k(n)+\mu\sum_{r\in\mathcal{N}_k\cap\mathcal{C}(k)}c_{k,r}\sum_{i=0}^{N_D-1}\frac{\varphi\left[d_{r,iD}(n)-\boldsymbol{u}^T_{r,i}(n)\boldsymbol{w}_k(n)\right]}{\left\|\boldsymbol{u}_{r,i}(n)\right\|^2}\boldsymbol{u}_{r,i}(n)\\&\quad+\mu\eta\sum_{l\in\mathcal{N}_k\setminus\mathcal{C}(k)}\gamma_{k,l}\left(\boldsymbol{w}_l(n)-\boldsymbol{w}_k(n)\right)+2\mu\sum_{m\in\mathcal{N}_k^-\cap\mathcal{C}(k)}b_{m,k}\left(\boldsymbol{w}^0_m-\boldsymbol{w}_k(n)\right)\end{aligned}\tag{6}$$

where $\mu$ is the step size, the step size of all nodes can be different, but for simplicity, the same value is chosen. By replacing the first three terms on the right-hand side of (6) with the intermediate estimate vector $\boldsymbol{\psi}_k(n+1)$, then updating formula (6) can be performed in the following two steps:

$$\begin{aligned}\boldsymbol{\psi}_k(n+1)&=\boldsymbol{w}_k(n)+\mu\sum_{r\in\mathcal{N}_k\cap\mathcal{C}(k)}c_{k,r}\sum_{i=0}^{N_D-1}\frac{\varphi\left[d_{r,iD}(n)-\boldsymbol{u}^T_{r,i}(n)\boldsymbol{w}_k(n)\right]}{\left\|\boldsymbol{u}_{r,i}(n)\right\|^2}\boldsymbol{u}_{r,i}(n)\\&\quad+\mu\eta\sum_{l\in\mathcal{N}_k\setminus\mathcal{C}(k)}\gamma_{k,l}\left(\boldsymbol{w}_l(n)-\boldsymbol{w}_k(n)\right)\\\boldsymbol{w}_k(n+1)&=\boldsymbol{\psi}_k(n+1)+2\mu\sum_{m\in\mathcal{N}_k^-\cap\mathcal{C}(k)}b_{m,k}\left(\boldsymbol{w}^0_m-\boldsymbol{w}_k(n)\right)\end{aligned}\tag{7}$$

Actually, $\boldsymbol{w}^0_m$ is unavailable for node *m* and can be replaced by the intermediate estimate $\boldsymbol{\psi}_m(n+1)$. Moreover, we can use $\boldsymbol{\psi}_k(n+1)$ to replace $\boldsymbol{w}_k(n)$ since the former contains more information than latter. Therefore, (7) can be rewritten as

$$
\begin{cases}
\boldsymbol{\psi}_k(n+1) = \boldsymbol{w}_k(n) + \mu \sum_{r \in \mathcal{N}_k \cap \mathcal{C}(k)} c_{k,r} \sum_{i=0}^{N_D-1} \frac{\varphi\left[d_{r,iD}(n) - \boldsymbol{u}_{r,i}^T(n)\boldsymbol{w}_k(n)\right]}{\left\|\boldsymbol{u}_{r,i}(n)\right\|^2} \boldsymbol{u}_{r,i}(n) \\
\qquad + \mu\eta \sum_{l \in \mathcal{N}_k \setminus \mathcal{C}(k)} \gamma_{k,l}\left(\boldsymbol{w}_l(n) - \boldsymbol{w}_k(n)\right) \\
\boldsymbol{w}_k(n+1) = \sum_{m \in \mathcal{N}_k \cap \mathcal{C}(k)} \alpha_{m,k} \boldsymbol{\psi}_m(n+1)
\end{cases} \tag{8}
$$

where $\alpha_{m,k} \triangleq 2\mu b_{m,k} (m \neq k)$ and $\alpha_{k,k} \triangleq 1 - 2\mu \sum_{m \in \mathcal{N}_k^- \cap \mathcal{C}(k)} b_{m,k}$ .

It can be seen from (8) that updating the intermediate estimate $\boldsymbol{\psi}_k(n+1)$ needs to receive data pairs $\left\{d_{r,iD}(n), \boldsymbol{u}_{r,i}(n)\right\}$ from nodes belong to the same cluster in the neighboring nodes of node *k*, which yields a considerable communication burden. To reduce the communication burden, we set $c_{k,r} = 0 (k \neq r)$ and $c_{k,k} = 1$ , i.e., node *k* only uses its own data for adaptive update, see [23] for the same operation. Thus, (8) can be further simplified as

$$
\begin{cases}
\boldsymbol{\psi}_k(n+1) = \boldsymbol{w}_k(n) + \mu \sum_{i=0}^{N_D-1} \frac{\varphi\left[e_{k,iD}(n)\right]}{\left\|\boldsymbol{u}_{k,i}(n)\right\|^2} \boldsymbol{u}_{k,i}(n) + \mu\eta \sum_{l \in \mathcal{N}_k \cap \mathcal{C}(k)} \gamma_{k,l}\left(\boldsymbol{w}_l(n) - \boldsymbol{w}_k(n)\right) \\
\boldsymbol{w}_k(n+1) = \sum_{m \in \mathcal{N}_k \cap \mathcal{C}(k)} \alpha_{m,k} \boldsymbol{\psi}_m(n+1)
\end{cases} \tag{9}
$$

where $e_{k,iD}(n) = d_{k,iD}(n) - \boldsymbol{u}_{k,i}^T(n)\boldsymbol{w}_k(n)$ , and

$$
\varphi[e_{k,iD}(n)] = \begin{cases} e_{k,iD}(n), & \left|e_{k,iD}(n)\right| < \xi_{k,iD}(n) \\ 0, & \text{other} \end{cases} \tag{10}
$$

***Remark***: In fact, the NSAF and APA are the extension of the NLMS algorithm in subband domain and time domain, respectively. From this point of view, the proposed MD-NMSAF algorithm can also be easily deduced from the update formula of MD-APM. The update formula of the MD-APM algorithm is given by [9]:

$$
\begin{cases}
\boldsymbol{\psi}_k(t+1) = \boldsymbol{w}_k(t) + \mu \underbrace{\boldsymbol{U}_k(t)\left(\varepsilon \boldsymbol{I} + \boldsymbol{U}_k^T(t)\boldsymbol{U}_k(t)\right)^{-1} \varphi\left[\boldsymbol{e}_k(t)\right]}_{\text{(a)}} \\
\qquad + \mu\eta \sum_{l \in \mathcal{N}_k \cap \mathcal{C}(k)} \gamma_{kl}\left(\boldsymbol{w}_l(t) - \boldsymbol{w}_k(t)\right) \\
\boldsymbol{w}_k(t+1) = \sum_{m \in \mathcal{N}_k \cap \mathcal{C}(k)} \alpha_{m,k} \boldsymbol{\psi}_m(t+1)
\end{cases}
$$

When the projection order *P* of MD-APA is reduced to 1, and the input signal and error signal in term (a) are replaced by the band-dependent input and error signals, respectively, i.e., the term (a) becomes $\sum_{i=0}^{N_D-1} \frac{\varphi\left[e_{k,iD}(n)\right]}{\left\|\boldsymbol{u}_{k,i}(n)\right\|^2} \boldsymbol{u}_{k,i}(n)$ , the proposed MD-NMSAF algorithm can be obtained, which also verifies the correctness of the derivation algorithm process in this section.

## 3. Performance analysis

To promote the performance analysis of the algorithm, the following commonly used assumptions are given:

*Assumption* 1: The input signal $\boldsymbol{u}_{k,i}(n)$ is temporally and spatially independent with zero-mean and positive definite auto-correlation matrix $\boldsymbol{R}_{k,i} = \mathrm{E}\{\boldsymbol{u}_{k,i}(n)\boldsymbol{u}_{k,i}^T(n)\}$ [5].

*Assumption* 2: $\tilde{\boldsymbol{w}}_k(n)$ is statistically independent of $\boldsymbol{u}_{k,i}(n)$ for all node *k* and $i = 0,\ldots,N_D - 1$.

*Assumption* 3: The Contaminated Gaussian (CG) model $v_k(t) = v_{g,k}(t) + v_{\eta,k}(t)$ is used as the measurement noise [12], $v_{g,k}(t)$ denotes the background noise, which is white Gaussian noise (WGN) with zero-mean and variance $\sigma_{g_k}^2$, $v_{\eta,k}(t) = b(t)\cdot c_k(t)$ is a Bernoulli Gaussian process, $b(t)$ is a Bernoulli process with probability mass function (PMF) $P(b(t)=1) = p_r$ and $P(b(t)=0) = 1 - p_r$, $c_k(t)$ is WGN with zero-mean and variance $\sigma_{c_k}^2$ ($\sigma_{c_k}^2 \gg \sigma_{g_k}^2$).

Using the update probability $P_{k,iD}(n) = P\left\{\left|e_{k,iD}(n)\right| < \xi_{k,iD}(n)\right\}$ [5], [12], the update formula (9) can be equivalently expressed as

$$\begin{aligned}\boldsymbol{w}_k(n+1) = \sum_{m\in\mathcal{N}_k\cap\mathcal{C}(k)} \alpha_{m,k} \Bigg[ & \boldsymbol{w}_m(n) + \mu \sum_{i=0}^{N_D-1} \frac{P_{m,iD}(n)\cdot e_{m,iD}(n)}{\left\|\boldsymbol{u}_{m,i}(n)\right\|^2} \boldsymbol{u}_{m,i}(n) \\ & + \mu\eta \sum_{l\in\mathcal{N}_m\setminus\mathcal{C}(m)} \gamma_{m,l}\left(\boldsymbol{w}_l(n) - \boldsymbol{w}_m(n)\right) \Bigg]\end{aligned} \tag{11}$$

The decimated subband reference signals can be rewritten as $d_{k,iD}(n) = \boldsymbol{u}_{k,i}^T(n)\boldsymbol{w}_k^* + v_{k,iD}(n)$, where $v_{k,iD}(n)$ stands for the decimated subband measurement noise of $v_k(t)$. The weight error vector of node *k* is defined as $\tilde{\boldsymbol{w}}_k(n) \triangleq \boldsymbol{w}_k^* - \boldsymbol{w}_k(n)$. Then, the estimation error of node *k* at time instant *n* on the *i*-th subband can be expressed as $e_{k,iD}(n) = \boldsymbol{u}_{k,i}^T(n)\tilde{\boldsymbol{w}}_k(n) + v_{g,k,iD}(n)$, where $v_{g,k,iD}(n)$ denote the decimated subband background noises of $v_{g,k}(t)$. Furthermore, (11) can be simplified as

$$\begin{aligned}\boldsymbol{w}_k(n+1) = \sum_{m\in\mathcal{N}_k\cap\mathcal{C}(k)} \alpha_{m,k} \Bigg[ & \boldsymbol{w}_m(n) + \mu\boldsymbol{B}_m(n)\tilde{\boldsymbol{w}}_m(n) \\ & + \mu\boldsymbol{T}_k(n) + \mu\eta \sum_{l\in\mathcal{N}_m\setminus\mathcal{C}(m)} \gamma_{m,l}\left(\boldsymbol{w}_l(n) - \boldsymbol{w}_m(n)\right) \Bigg]\end{aligned} \tag{12}$$

where $\boldsymbol{B}_k(n) = \sum_{i=0}^{N_D-1} P_{k,iD}(n)\boldsymbol{A}_{k,i}(n)$, $\boldsymbol{A}_{k,i}(n) = \dfrac{\boldsymbol{u}_{k,i}(n)\boldsymbol{u}_{k,i}^T(n)}{\left\|\boldsymbol{u}_{k,i}(n)\right\|^2}$, $\boldsymbol{T}_k(n) = \sum_{i=0}^{N_D-1} P_{k,iD}(n) v_{g,k,iD}(n)\boldsymbol{q}_{k,i}(n)$, and $\boldsymbol{q}_{k,i}(n) = \dfrac{\boldsymbol{u}_{k,i}(n)}{\left\|\boldsymbol{u}_{k,i}(n)\right\|^2}$.

To facilitate the analysis, some global quantities are defined as follows

$$\boldsymbol{w}^* \triangleq \text{col}[\boldsymbol{w}_1^*,\ldots,\boldsymbol{w}_N^*]$$
$$\boldsymbol{w}(n) \triangleq \text{col}\left[\boldsymbol{w}_1(n),\ldots,\boldsymbol{w}_N(n)\right]$$
$$\boldsymbol{\psi}(n) \triangleq \text{col}\left[\boldsymbol{\psi}_1(n),\ldots,\boldsymbol{\psi}_N(n)\right]$$
$$\boldsymbol{B}(n) \triangleq \text{diag}\left\{\boldsymbol{B}_1(n),\ldots,\boldsymbol{B}_N(n)\right\}$$
$$\boldsymbol{T}(n) \triangleq \text{col}\left\{\boldsymbol{T}_1(n),\ldots,\boldsymbol{T}_N(n)\right\}$$

According to the above global quantities, the global form of (12) is given by

$$\boldsymbol{w}(n+1) = \boldsymbol{G}\boldsymbol{w}(n) + \mu\boldsymbol{G}\boldsymbol{B}(n)\tilde{\boldsymbol{w}}(n) + \mu\boldsymbol{G}\boldsymbol{T}(n) - \mu\eta\boldsymbol{G}\boldsymbol{Q}\boldsymbol{w}(n) \tag{13}$$

where $\boldsymbol{G} = \boldsymbol{C}^T \otimes \boldsymbol{I}_M$, $[\boldsymbol{C}]_{m,k} = \alpha_{m,k}$, $\boldsymbol{Q} = \boldsymbol{I}_{NM} - \boldsymbol{P} \otimes \boldsymbol{I}_M$, $[\boldsymbol{P}]_{k,m} = \gamma_{k,m}$ and $\tilde{\boldsymbol{w}}(n) \triangleq \boldsymbol{w}^* - \boldsymbol{w}(n)$ is the network weight error vector.

Combining the network weight error vector $\tilde{\boldsymbol{w}}(n)$ and (13), we can obtain

$$\tilde{\boldsymbol{w}}(n+1) = \boldsymbol{G}\left[\boldsymbol{I}_{NM} - \mu\boldsymbol{Z}(n)\right]\tilde{\boldsymbol{w}}(n) - \mu\boldsymbol{G}\boldsymbol{T}(n) + \boldsymbol{\zeta} \tag{14}$$

where $\boldsymbol{Z}(n) = \boldsymbol{B}(n) + \eta\boldsymbol{Q}$ and $\boldsymbol{\zeta} = \mu\eta\boldsymbol{G}\boldsymbol{Q}\boldsymbol{w}^*$.

## 3.1 Mean stability

In this subsection, the mean stability of the proposed MD-NMSAF algorithm is analyzed, and the stable step size range is also given. Taking the expected values on both sides of (14), we have

$$\mathrm{E}\{\tilde{\boldsymbol{w}}(n+1)\} = \boldsymbol{G}\left[\boldsymbol{I}_{NM} - \mu\mathrm{E}\{\boldsymbol{Z}(n)\}\right]\mathrm{E}\{\tilde{\boldsymbol{w}}(n)\} + \boldsymbol{\zeta} \tag{15}$$

Now, it is only necessary to ensure that the spectral radius of $\boldsymbol{G}\left[\boldsymbol{I}_{NM} - \mu\mathrm{E}\{\boldsymbol{Z}(n)\}\right]$ is within the unit circle to make (15) stable, i.e.,

$$\rho\left(\boldsymbol{G}\left[\boldsymbol{I}_{NM} - \mu\mathrm{E}\{\boldsymbol{Z}(n)\}\right]\right) < 1 \tag{16}$$

According to the properties of the matrix spectral radius and norm [4], yields:

$$\begin{aligned}\rho\left(\boldsymbol{G}\left[\boldsymbol{I}_{NM} - \mu\mathrm{E}\{\boldsymbol{Z}(n)\}\right]\right) &\le \left\|\boldsymbol{G}\left[\boldsymbol{I}_{NM} - \mu\mathrm{E}\{\boldsymbol{Z}(n)\}\right]\right\|_{b,\infty} \\ &\le \|\boldsymbol{G}\|_{b,\infty}\left\|\boldsymbol{I}_{NM} - \mu\mathrm{E}\{\boldsymbol{Z}(n)\}\right\|_{b,\infty}\end{aligned} \tag{17}$$

where $\|\cdot\|_{b,\infty}$ is the block-maximum-norm [4]. Since the combination weight matrix satisfies $\boldsymbol{C}^T\boldsymbol{1} = \boldsymbol{1}$, where $\boldsymbol{1}$ is the column vector whose elements are all 1, $\|\boldsymbol{G}\|_{b,\infty} = \left\|\boldsymbol{C}^T\right\|_\infty = 1$ holds, and the following relation is established

$$\begin{aligned}&\|\boldsymbol{G}\|_{b,\infty}\left\|\boldsymbol{I}_{NM} - \mu\mathrm{E}\left\{\boldsymbol{Z}(n)\right\}\right\|_{b,\infty} \\ &= \left\|\boldsymbol{I}_{NM} - \mu\mathrm{E}\left\{\boldsymbol{Z}(n)\right\}\right\|_{b,\infty} \\ &\le \left\|\boldsymbol{I}_{NM} - \mu\mathrm{E}\{\boldsymbol{B}(n)\} - \mu\eta\boldsymbol{I}_{NM}\right\|_{b,\infty} + \mu\eta\left\|\boldsymbol{P}\otimes\boldsymbol{I}_M\right\|_{b,\infty} \\ &\overset{\text{(b)}}{=} \rho\left(\boldsymbol{I}_{NM} - \mu\mathrm{E}\left\{\boldsymbol{B}(n)\right\} - \mu\eta\boldsymbol{I}_{NM}\right) + \mu\eta\end{aligned} \tag{18}$$

where (b) is obtained due to $\|\boldsymbol{P}\otimes\boldsymbol{I}_M\|_{b,\infty}=1$ [4].

Then, the mean stability condition of the MD-NMSAF algorithm becomes

$$\rho\left(\boldsymbol{I}_{NM}-\mu\mathrm{E}\{\boldsymbol{B}(n)\}-\mu\eta\boldsymbol{I}_{NM}\right)+\mu\eta<1 \tag{19}$$

Finally, the range of step size is given by

$$0<\mu<\frac{2}{\max\limits_{k}\{\lambda_{\max}(\mathrm{E}\{\boldsymbol{B}_k\})\}+2\eta} \tag{20}$$

where $\lambda_{\max}(\cdot)$ represents the largest eigenvalue of the matrix.

## 3.2 Mean square stability

By using Σ-weighted Euclidean norm on both sides of (14) and taking the expected values, we obtain

$$\begin{aligned}\mathrm{E}\left\{\|\tilde{\boldsymbol{w}}(n+1)\|_{\Sigma}^2\right\}&=\mathrm{E}\left\{\|\tilde{\boldsymbol{w}}(n)\|_{\Sigma_1}^2\right\}+\mu^2\mathrm{E}\left\{\boldsymbol{T}^T(n)\boldsymbol{G}^T\Sigma\boldsymbol{G}\boldsymbol{T}(n)\right\}\\&+2\mathrm{E}\left\{\tilde{\boldsymbol{w}}^T(n)\right\}\left(\left[\boldsymbol{I}-\mu\mathrm{E}\left\{\boldsymbol{Z}^T(n)\right\}\right]\boldsymbol{G}^T\right)\Sigma\boldsymbol{\zeta}+\boldsymbol{\zeta}^T\Sigma\boldsymbol{\zeta}\end{aligned} \tag{21}$$

where $\mathrm{E}\left\{\|\tilde{\boldsymbol{w}}(n+1)\|_{\Sigma}^2\right\}=\tilde{\boldsymbol{w}}^T(n+1)\Sigma\tilde{\boldsymbol{w}}(n+1)$ , and

$$\begin{aligned}\Sigma_1=&\,\boldsymbol{G}^T\Sigma\boldsymbol{G}-\mu\boldsymbol{G}^T\Sigma\boldsymbol{G}\mathrm{E}\{\boldsymbol{Z}(n)\}-\mu\mathrm{E}\left\{\boldsymbol{Z}^T(n)\right\}\boldsymbol{G}^T\Sigma\boldsymbol{G}\\&+\mu^2\mathrm{E}\left\{\boldsymbol{Z}^T(n)\right\}\boldsymbol{G}^T\Sigma\boldsymbol{G}\mathrm{E}\left\{\boldsymbol{Z}(n)\right\}\end{aligned} \tag{22}$$

Let $\mathrm{vec}(\cdot)$ and $\otimes$ denote the vectorization operation and Kronecker product operator, respectively. Using the vectorization operation on both sides of (22) and the property $\mathrm{vec}(X\Sigma Y)=(Y^T\otimes X)\,\mathrm{vec}(\Sigma)$ , yields

$$\mathrm{vec}(\Sigma_1)=\boldsymbol{K}(\boldsymbol{G}^T\otimes\boldsymbol{G}^T)\mathrm{vec}(\Sigma) \tag{23}$$

where $\boldsymbol{K}=\boldsymbol{I}_{N^2M^2}-\mu\boldsymbol{\Lambda}+\mu^2\boldsymbol{\Gamma}$ , $\boldsymbol{\Gamma}$=$\mathrm{E}\left\{\boldsymbol{Z}^T(n)\right\}\otimes\mathrm{E}\left\{\boldsymbol{Z}^T(n)\right\}$ and $\boldsymbol{\Lambda}=\left(\mathrm{E}\left\{\boldsymbol{Z}^T(n)\right\}\otimes\boldsymbol{I}_{NM}\right)+\left(\boldsymbol{I}_{NM}\otimes\mathrm{E}\left\{\boldsymbol{Z}^T(n)\right\}\right)$.

It is known from [24] that $\mathrm{E}\{\boldsymbol{T}(n)\boldsymbol{T}^T(n)\}$ and $\mathrm{E}\{\boldsymbol{B}(n)\}$ are finite values, so the terms $\mathrm{E}\left\{\boldsymbol{T}^T(n)\boldsymbol{G}^T\Sigma\boldsymbol{G}\boldsymbol{T}(n)\right\}$=$\mathrm{Tr}\left(\boldsymbol{G}^T\Sigma\boldsymbol{G}\mathrm{E}\left\{\boldsymbol{T}(n)\boldsymbol{T}^T(n)\right\}\right)$ and $\mathrm{E}\left\{\boldsymbol{Z}^T(n)\right\}$ in the right side of (21) are finite values. When the MD-NMSAF algorithm satisfy the mean stability, $\mathrm{E}\left\{\tilde{\boldsymbol{w}}^T(n)\right\}$ will converge to a finite value, so the last three terms on the right side of (21) are all finite values. To ensure the mean square stability of the MD-NMSAF, the spectral radius of $\boldsymbol{K}\left(\boldsymbol{G}^T\otimes\boldsymbol{G}^T\right)$ must satisfy $\rho(\boldsymbol{K}\left(\boldsymbol{G}^T\otimes\boldsymbol{G}^T\right))\le 1$.

According to the properties of the spectral radius and matrix norm, we have the following relation

$$\rho\left(\boldsymbol{K}(\boldsymbol{G}^T\otimes\boldsymbol{G}^T)\right)\le\left\|\boldsymbol{K}\left(\boldsymbol{G}^T\otimes\boldsymbol{G}^T\right)\right\|_{b,\infty}\le\|\boldsymbol{K}\|_{b,\infty}\left\|\boldsymbol{G}^T\otimes\boldsymbol{G}^T\right\|_{b,\infty} \tag{24}$$

Since $\left\| \boldsymbol{G}^T \otimes \boldsymbol{G}^T \right\|_{b,\infty} = 1$, and $\boldsymbol{K}$ is a block diagonal Hermite matrix whose block maximum norm is equal to its spectral radius. Inequation (24) can be rewritten as

$$\rho\left(\boldsymbol{K}(\boldsymbol{G}^T \otimes \boldsymbol{G}^T)\right) \le \max\left|\lambda\left(\boldsymbol{K}\right)\right| \le 1 \tag{25}$$

Following the same argument used in [25], the mean square stability condition of the MD-NMSAF algorithm is given by

$$0 < \mu < \min\left\{\frac{1}{\lambda_{\max}(\boldsymbol{\Lambda}^{-1}\boldsymbol{\Gamma})}, \frac{1}{\max\left(\lambda(\boldsymbol{\Phi}) \in \Re^{+}\right)}\right\} \tag{26}$$

where $\boldsymbol{\Phi} = \begin{pmatrix} \frac{1}{2}\boldsymbol{\Lambda} & -\frac{1}{2}\boldsymbol{\Gamma} \\ \boldsymbol{I}_{N^2M^2} & \boldsymbol{0} \end{pmatrix}$.

## 3.3 Transient and Steady-state network MSD

Defining the auto-correlation matrix $\mathcal{W}(n) \triangleq \mathrm{E}\{\tilde{\boldsymbol{w}}(n)\tilde{\boldsymbol{w}}^T(n)\}$ of the network weight error vector $\tilde{\boldsymbol{w}}(n)$. And according to (14), we have

$$\begin{aligned} \mathcal{W}(n+1) = {} & \boldsymbol{G}\mathrm{E}\left\{[\boldsymbol{I} - \mu\boldsymbol{Z}(n)]\tilde{\boldsymbol{w}}(n)\tilde{\boldsymbol{w}}^T(n)[\boldsymbol{I} - \mu\boldsymbol{Z}(n)]^T\right\}\boldsymbol{G}^T \\ & +\boldsymbol{G}\mathrm{E}\left\{[\boldsymbol{I} - \mu\boldsymbol{Z}(n)]\tilde{\boldsymbol{w}}(n)\right\}\boldsymbol{\zeta}^T \\ & +\mu^2\boldsymbol{G}\mathrm{E}\left\{\boldsymbol{T}(n)\boldsymbol{T}^T(n)\right\}\boldsymbol{G}^T \\ & +\boldsymbol{\zeta}\mathrm{E}\left\{\tilde{\boldsymbol{w}}^T(n)[\boldsymbol{I} - \mu\boldsymbol{Z}(n)]^T\right\}\boldsymbol{G}^T + \boldsymbol{\zeta}\boldsymbol{\zeta}^T \end{aligned} \tag{27}$$

The network MSD at time instant $n$ is defined as

$$\mathrm{MSD}(n) = \frac{1}{N}\mathrm{Tr}\left\{\mathcal{W}(n)\right\} \tag{28}$$

Vectoring both sides of (27) yields

$$\begin{aligned} \mathrm{vec}\left(\mathcal{W}(n+1)\right) = {} & \mathcal{F}(n)\mathrm{vec}\left(\mathcal{W}(n)\right) \\ & +(\boldsymbol{I}\otimes\boldsymbol{G})\mathrm{vec}\left(\mathrm{E}\{\tilde{\boldsymbol{w}}(n)\}\boldsymbol{\zeta}^T\right) - \mu(\boldsymbol{I}\otimes\boldsymbol{G})\mathrm{vec}\left(\mathrm{E}\left\{\boldsymbol{Z}(n)\tilde{\boldsymbol{w}}(n)\right\}\boldsymbol{\zeta}^T\right) \\ & +\mu^2(\boldsymbol{G}\otimes\boldsymbol{G})\mathrm{vec}\left(\mathrm{E}\left\{\boldsymbol{T}(n)\boldsymbol{T}^T(n)\right\}\right) \\ & +(\boldsymbol{G}\otimes\boldsymbol{I})\mathrm{vec}\left(\boldsymbol{\zeta}\mathrm{E}\{\tilde{\boldsymbol{w}}^T(n)\}\right) - \mu(\boldsymbol{G}\otimes\boldsymbol{I})\mathrm{vec}\left(\boldsymbol{\zeta}\mathrm{E}\left\{\tilde{\boldsymbol{w}}^T(n)\boldsymbol{Z}^T(n)\right\}\right) \\ & +\mathrm{vec}\left(\boldsymbol{\zeta}\boldsymbol{\zeta}^T\right) \end{aligned} \tag{29}$$

where

$$\begin{aligned} \mathcal{F}(n) = {} & (\boldsymbol{G}\otimes\boldsymbol{G})\left[\boldsymbol{I}_{N^2M^2} - \mu\left(\boldsymbol{I}_{NM} \otimes \mathrm{E}\{\boldsymbol{Z}(n)\}\right)\right. \\ & \left. -\mu\left(\mathrm{E}\{\boldsymbol{Z}(n)\} \otimes \boldsymbol{I}_{NM}\right) + \mu^2\mathrm{E}\left\{\boldsymbol{Z}(n) \otimes \boldsymbol{Z}(n)\right\}\right] \end{aligned} \tag{30}$$

According to $\mathrm{Tr}(\boldsymbol{S}_1\boldsymbol{S}_2) = \mathrm{vec}(\boldsymbol{S}_1^T)^T\,\mathrm{vec}(\boldsymbol{S}_2)$, the network MSD at time instant $n$ is given by

$$\mathrm{MSD}(n)=\frac{1}{N}\mathrm{vec}(\boldsymbol{I}_{NM})^T\,\mathrm{vec}\big(\mathcal{W}(n)\big) \tag{31}$$

When the MD-NMSAF algorithm converges to the steady-state, (29) can be rewritten as

$$\begin{aligned}\mathrm{vec}(\mathcal{W}(\infty))=&\left(\boldsymbol{I}_{N^2M^2}-\mathcal{F}(\infty)\right)^{-1}\\&\cdot\Big[(\boldsymbol{I}\otimes\boldsymbol{G})\mathrm{vec}\big(\mathrm{E}\{\tilde{\boldsymbol{w}}(\infty)\}\boldsymbol{\zeta}^T\big)-\mu(\boldsymbol{I}\otimes\boldsymbol{G})\mathrm{vec}\big(\mathrm{E}\{\boldsymbol{Z}(\infty)\tilde{\boldsymbol{w}}(\infty)\}\boldsymbol{\zeta}^T\big)\\&+\mu^2(\boldsymbol{G}\otimes\boldsymbol{G})\mathrm{vec}\big(\mathrm{E}\{\boldsymbol{T}(\infty)\boldsymbol{T}^T(\infty)\}\big)\\&+(\boldsymbol{G}\otimes\boldsymbol{I})\mathrm{vec}\big(\boldsymbol{\zeta}\mathrm{E}\{\tilde{\boldsymbol{w}}^T(\infty)\}\big)-\mu(\boldsymbol{G}\otimes\boldsymbol{I})\mathrm{vec}\big(\boldsymbol{\zeta}\mathrm{E}\{\tilde{\boldsymbol{w}}^T(\infty)\boldsymbol{Z}^T(\infty)\}\big)\\&+\mathrm{vec}\big(\boldsymbol{\zeta}\boldsymbol{\zeta}^T\big)\Big]\end{aligned} \tag{32}$$

The steady-state network MSD can be expressed as

$$\begin{aligned}\mathrm{MSD}(\infty)=&\frac{1}{N}\mathrm{vec}(\boldsymbol{I}_{NM})^T\left(\boldsymbol{I}_{N^2M^2}-\mathcal{F}(\infty)\right)^{-1}\\&\cdot\Big[(\boldsymbol{I}\otimes\boldsymbol{G})\mathrm{vec}\big(\mathrm{E}\{\tilde{\boldsymbol{w}}(\infty)\}\boldsymbol{\zeta}^T\big)-\mu(\boldsymbol{I}\otimes\boldsymbol{G})\mathrm{vec}\big(\mathrm{E}\{\boldsymbol{Z}(\infty)\tilde{\boldsymbol{w}}(\infty)\}\boldsymbol{\zeta}^T\big)\\&+\mu^2(\boldsymbol{G}\otimes\boldsymbol{G})\mathrm{vec}\big(\mathrm{E}\{\boldsymbol{T}(\infty)\boldsymbol{T}^T(\infty)\}\big)\\&+(\boldsymbol{G}\otimes\boldsymbol{I})\mathrm{vec}\big(\boldsymbol{\zeta}\mathrm{E}\{\tilde{\boldsymbol{w}}^T(\infty)\}\big)-\mu(\boldsymbol{G}\otimes\boldsymbol{I})\mathrm{vec}\big(\boldsymbol{\zeta}\mathrm{E}\{\tilde{\boldsymbol{w}}^T(\infty)\boldsymbol{Z}^T(\infty)\}\big)\\&+\mathrm{vec}\big(\boldsymbol{\zeta}\boldsymbol{\zeta}^T\big)\Big]\end{aligned} \tag{33}$$

where $\mathrm{E}\{\tilde{\boldsymbol{w}}(\infty)\}=\left(\boldsymbol{I}_{NM}-\boldsymbol{G}\left[\boldsymbol{I}_{NM}-\mu\mathrm{E}\{\boldsymbol{Z}(\infty)\}\right]\right)^{-1}\boldsymbol{\zeta}$ is obtained from (15) when $n\to\infty$, and

$$\begin{aligned}\mathcal{F}(\infty)=(\boldsymbol{G}\otimes\boldsymbol{G})\Big[&\boldsymbol{I}_{N^2M^2}-\mu\big(\boldsymbol{I}_{NM}\otimes\mathrm{E}\{\boldsymbol{Z}(\infty)\}\big)\\&-\mu\big(\mathrm{E}\{\boldsymbol{Z}(\infty)\}\otimes\boldsymbol{I}_{NM}\big)+\mu^2\mathrm{E}\{\boldsymbol{Z}(\infty)\otimes\boldsymbol{Z}(\infty)\}\Big]\end{aligned} \tag{34}$$

## 3.4 Computational Complexity

TABLE I Computational complexity

| Algorithms | Multiplications | Additions | DMI |
|---|---|---|---|
| MD-LMS | $\sum_{k=1}^{N}\left[(m_k+n_k+1)M+1\right]$ | $\sum_{k=1}^{N}\left[(m_k+n_k)M\right]$ | 0 |
| MD-APA | $\sum_{k=1}^{N}\left[(P^2+2P+m_k+n_k)M+P^3+P^2\right]$ | $\sum_{k=1}^{N}\left[(P^2+2P+2m_k+n_k-1)M+P^3\right]$ | $O(P^3)$ |
| MD-APM | $\sum_{k=1}^{N}\left[(P^2+2P+m_k+n_k)M+P^3+P^2\right]$ | $\sum_{k=1}^{N}\left[(P^2+2P+2m_k+n_k-1)M+P^3\right]$ | $O(P^3)$ |
| MD-APMCC | $\sum_{k=1}^{N}\left[(P^2+2P+m_k+n_k)M+P^3+P^2+6P\right]$ | $\sum_{k=1}^{N}\left[(P^2+2P+2m_k+n_k-1)M+P^3\right]$ | $O(P^3)$ |
| MD-NMSAF | $\sum_{k=1}^{N}\left[(m_k+n_k+N_D+2)M+2\right]$ | $\sum_{k=1}^{N}\left[(m_k+n_k+2N_D-1)M-N_D\right]$ | 0 |

Table I summarizes the computational complexity of the proposed MD-NMSAF algorithm, MD-LMS algorithm and some AP-based multitask diffusion algorithms at each iteration in terms of addition/minus, multiplication, and direct matrix inversion (DMI) operations. $N$ is the number of node in the network; $m_k$ and $n_k$ denote the number of node in the sets $\mathcal{N}_k/\mathcal{C}(k)$ and $\mathcal{N}_k\cap\mathcal{C}(k)$, respectively; $M$ is the length of filter; $N_D$ is the number of subband; $P$ is the projection order. Although the AP based multitask diffusion adaptive algorithms significantly improve the convergence speed of the MD-LMS algorithm under correlated input signal. unfortunately, the computational complexity has also increased significantly. Compared with the AP-type multitask diffusion algorithms, the MD-NMSAF algorithm proposed in this paper reduces the computational complexity a lot, and is close to the MD-LMS algorithm when the subband number is small.

## 4. Simulation results

In this section, two multitask networks with different topologies as shown in Fig.2 are used to verify the performance of the proposed MD-NMSAF algorithm. The multitask network with node $N$=7 is used to verify the validity of the theoretical analysis of the proposed MD-NMSAF algorithm. The multitask network with node $N$=15 is used to demonstrate the performance advantages of the MD-NMSAF algorithm over some existing multitask algorithms. As shown in Fig.2, the nodes in different color regions (clusters) in the topology correspond to different estimation tasks (target parameter vectors). The target parameter vectors for the three clusters can be generated by $\boldsymbol{w}_{C_l}^* = \boldsymbol{w}^* + \hbar_{C_l}\boldsymbol{w}^*$, $C_l \in \{C_1, C_2, C_3\}$, where $\hbar_{C_1} = 0.025$, $\hbar_{C_2} = -0.05$, $\hbar_{C_1} = 0.075$, and $\boldsymbol{w}^* = \text{rand}(M,1)$. The combination weight coefficients $\gamma_{k,l}$ in the multitask algorithms are selected as $\gamma_{k,l} = 1/|\mathcal{N}_k \setminus \mathcal{C}(k)|$ for $l \in \mathcal{N}_k \setminus \mathcal{C}(k)$, where $|\cdot|$ denotes the cardinality of a set, and $\alpha_{m,k}$ follow the Uniform rule in [26]. The network MSD is used as the performance measure of algorithms, which is defined as follows

$$\text{MSD}(n) = \frac{1}{N}\sum_{k=1}^{N}\text{E}\left\{\left\|\boldsymbol{w}_k^* - \boldsymbol{w}_k(n)\right\|_2^2\right\} \tag{35}$$

Three different input signals are used for our simulations, and described as follows:

(1) Gaussian input signal: The input signal of each node $k$ is a white Gaussian process with zero-mean and variance $\sigma_{\delta_k}^2$;

(2) AR (1) input signal: The input signal is produced by the first-order auto-regressive system $u_k(t)=\beta_1 u_k(t-1)+\delta_k(t)$, where $\beta_1$ is the correlation coefficient, $\delta_k(t)$ is white Gaussian process with zero-mean and variance $\sigma_{\delta_k}^2$.

(3) AR (2) input signal: The input signal is produced by the second-order auto-regressive system $u_k(t)=\beta_2 u_k(t-1)+\beta_3 u_k(t-2)+\delta_k(t)$, where $\beta_2$ and $\beta_3$ are the correlation coefficients.

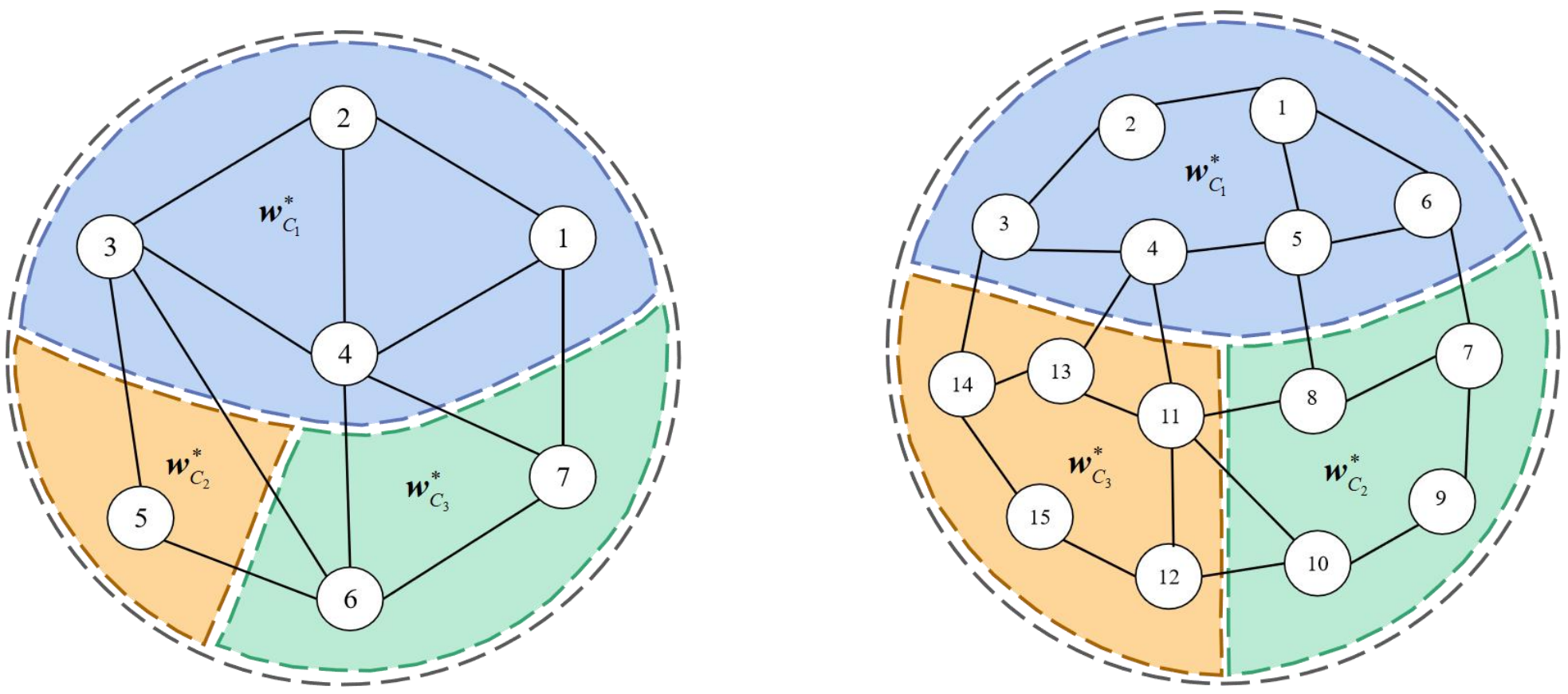

(a) (b)

Fig. 2 Multitask network topology. (a) $N = 7$ . (b) $N = 15$ .

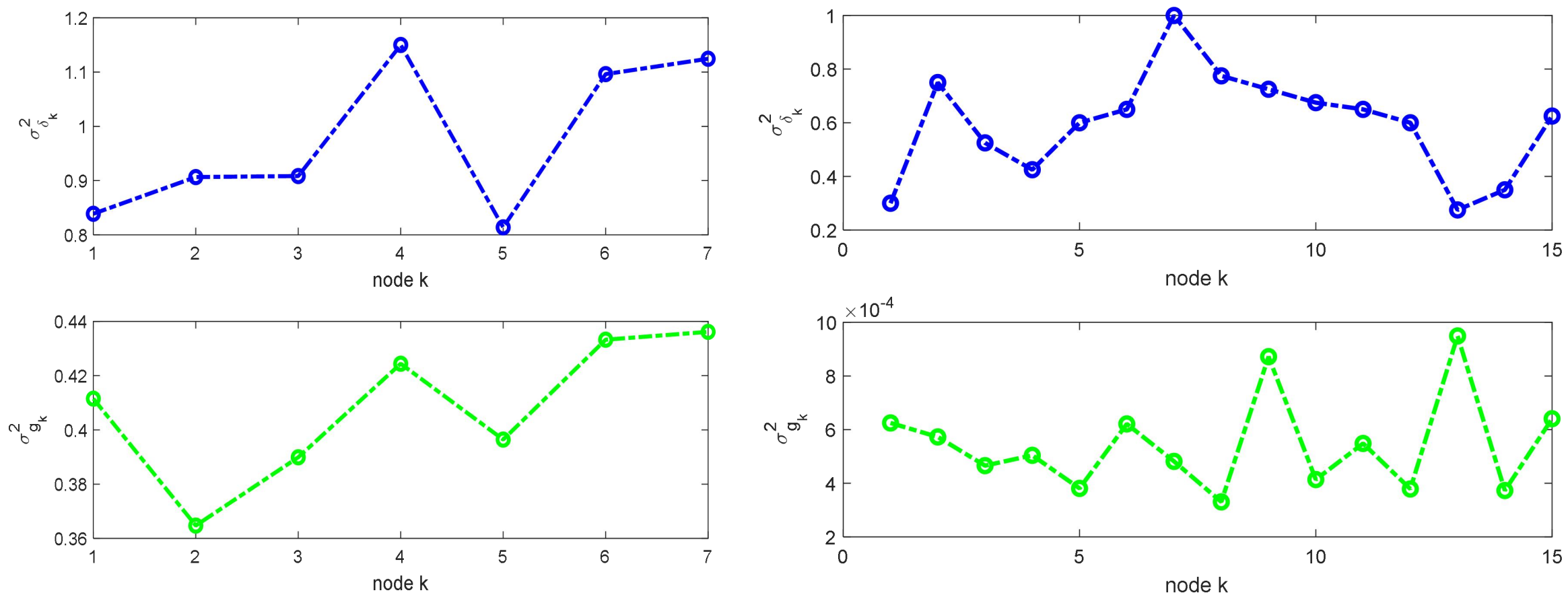

(a) (b)

Fig. 3 $\sigma^2_{\delta_k}$ and $\sigma^2_{g_k}$ of all nodes in multitask networks. (a) $N = 7$ . (b) $N = 15$ .

## 4.1 Verification of the theoretical analysis results

### 4.1.1 Steady-state performance

In this part, the correctness of the steady-state analysis result of the proposed MD-NMSAF algorithm is verified under the multitask network with $N=7$ nodes shown in Fig. 2(a). The filter order is $M=8$ . The parameter settings of three different input signals are as follows: $\beta_1 = 0.95$ , $\beta_2 = 0.1$ , $\beta_3 = 0.8$ , $\sigma^2_{\delta_k}$ of all nodes are shown in Fig.3 (a). The parameters of CG noise are given by: $p_r = 0.001$ , $\sigma^2_{c_k} = 10^3 \sigma^2_{g_k}$ , and the background noise variance $\sigma^2_{g_k}$ of all node are shown in Fig.3(a). The simulated steady-state MSD values are obtained by taking the average of the last 100 iterations of the total 50000 iterations.

In Fig.4, the steady-state network MSD of the MD-NMSAF algorithm versus step size and subband number are presented, and we can see from it that the simulated values are matched well with the theoretical analysis results, which confirms the validity of the steady-state analysis result (33) derived in Section 3.3. In addition, the conclusions under three different inputs are similar, that is, the smaller the number of subbands, the lower the steady-state network MSD of the MD-NMSAF. And the steady-state network

MSD gradually increases with the increase of step size.

#### 4.1.2 Transient performance

In this subsection, our purpose is to verify the validity of the transient performance analysis result, and study the influence of the number of subbands and step size on the proposed MD-NMSAF algorithm. Unless otherwise specified, the multitask network and parameters selection are the same as those in Section 4.1.1. Fig.5 shows the theoretical and simulated transient network MSD curves of the MD-NMSAF algorithm under different input signals, where the number of subbands $N_D$ is set to 2, 4, 8, and the step size is fixed to $\mu = 0.005$. Moreover, the theoretical and simulated transient network MSD curves under different step sizes are also provided in Fig.6, where the step size is set to 0.01, 0.02, 0.03, and the number of subbands $N_D$ is fixed to 8.

As displayed in Fig.5 and Fig.6, the theoretical transient network MSD curves computed by (31) are in good agreement with the simulation curves under different input signals, subband numbers and step sizes. Fig.5 demonstrates that the smaller the number of subbands, the lower the steady-state network MSD of MD-NMSAF, but the slower the convergence speed also. From Fig.6, one can see that the convergence speed of MD-NMSAF is gradually increased as the step size increases, but the steady-state network MSD is also increased.

#### 4.1.3 Stable step size range

In this simulation, the validity of the upper bound of the stable step size range calculated in Section 3 is verified under different input signals. The number of subbands $N_D$ of the MD-NMSAF algorithm is set to 4. The green dotted lines in Fig.7 depict the upper bound of the stable step size range (26), and the blue and red curves are the simulated and theoretical steady-state network MSD values, respectively. Simulation results show that when the step size of MD-NMSAF is near the upper bound of the stable step size range, the algorithm appears divergent, which verifies the accuracy of the stable step size bound.

### 4.2 Performance comparison

In this subsection, a multitask network with nodes $N = 15$ as shown in Fig.2 (b) is used to test the convergence and tracking performances of the MD-NMSAF algorithm with those of some existing multitask algorithms. The filter order is $M = 16$. The parameter settings of three different input signals are as follows: $\beta_1 = 0.9$, $\beta_2 = 0.1$ and $\beta_3 = 0.8$, $\sigma^2_{\delta_k}$ of all nodes are given by Fig.3 (b). The parameters of CG noise are as follows: $p_r = 0.01$, $\sigma^2_{c_k} = 10^4 \sigma^2_{g_k}$, the background noise variance $\sigma^2_{g_k}$ of all nodes are shown in Fig.3(b). To maintain fairness, the parameters of all algorithms are carefully selected to achieve the same initial convergence rate, as provided in Table II. It can be seen from Fig.8 that the convergence performances of the MD-APM and MD-NMSAF algorithms are almost the same under the white Gaussian inputs, while the MDAPA performs the worst since its not robust to impulsive noise. For AR(1) input signal, the MD-NMSAF algorithm has the lowest steady-state network MSD. And for AR(2) input signal, the convergence speed of MD-NMSAF is much faster than the MD-APM algorithm, which is attributed to the outstanding decorrelation ability of the SAF.

The tracking performance of algorithm is particularly important for time-varying systems, so we will checkout the tracking performance of the proposed MD-NMSAF algorithm through computer simulations. In this experiment, the unknown parameter vectors of all clusters change from $\boldsymbol{w}_{C_l}^{*}$ to $-\boldsymbol{w}_{C_l}^{*}$ at the middle iteration. The parameter settings of all algorithms are the same as those in Fig.8. As shown in Fig.9 (a), the tracking speeds of all algorithms are the same for white Gaussian inputs, but the steady-state network MSD of the MD-APM and MD-NMSAF algorithms are the lowest. On the contrary, the proposed MD-NMSAF algorithm can simultaneously provide the fastest tracking speed and the smallest steady-state error under correlated inputs and impulsive noise, which is due to the strong decorrelation ability of the SAF and the robustness of M-estimate function.

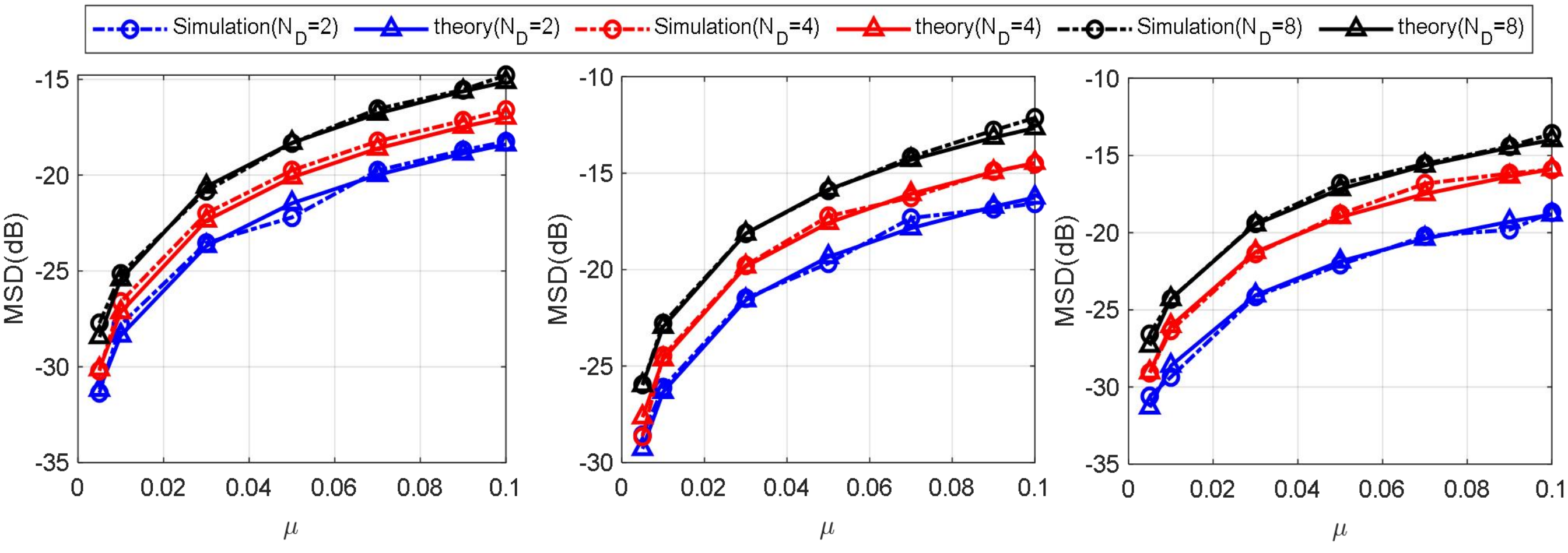

(a) (b) (c)

Fig. 4 Theoretical and simulated steady-state network MSD versus step size and subband number.

(a) Gaussian input signal. (b) AR (1) input signal. (c) AR (2) input signal.

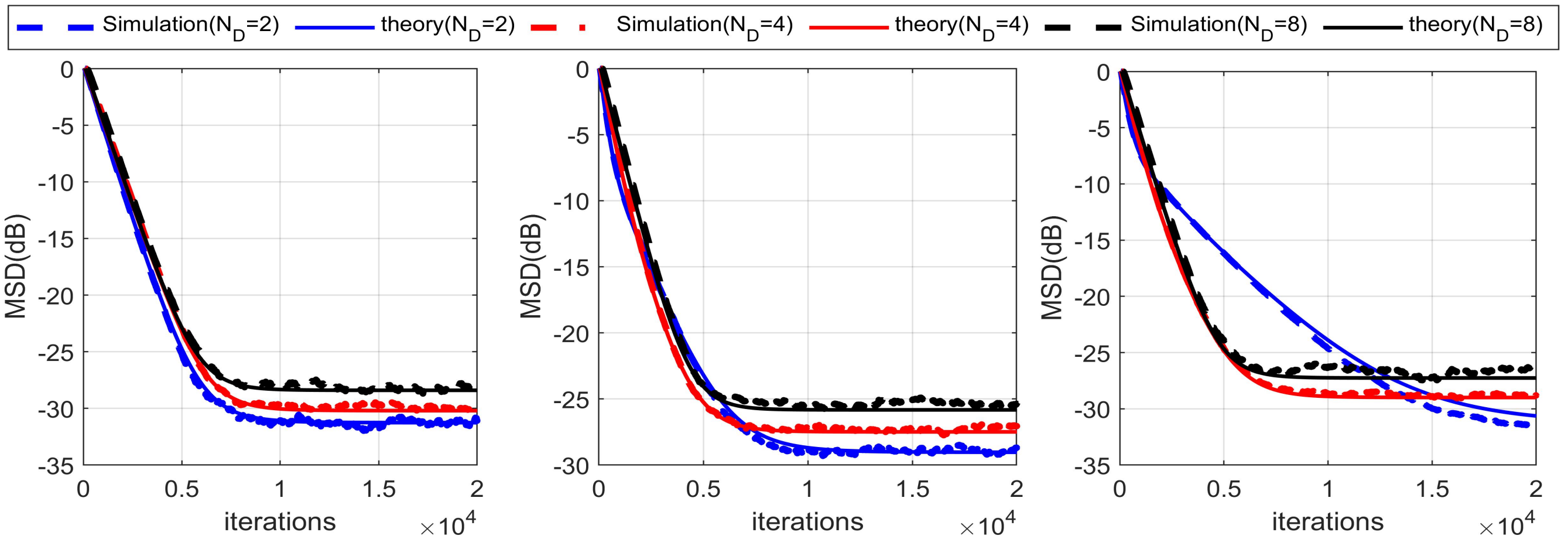

(a) (b) (c)

Fig. 5 Theoretical and simulated transient network MSD curves under different subband numbers. The step size is fixed to $\mu = 0.005$.

(a) Gaussian input signal. (b) AR (1) input signal. (c) AR (2) input signal.

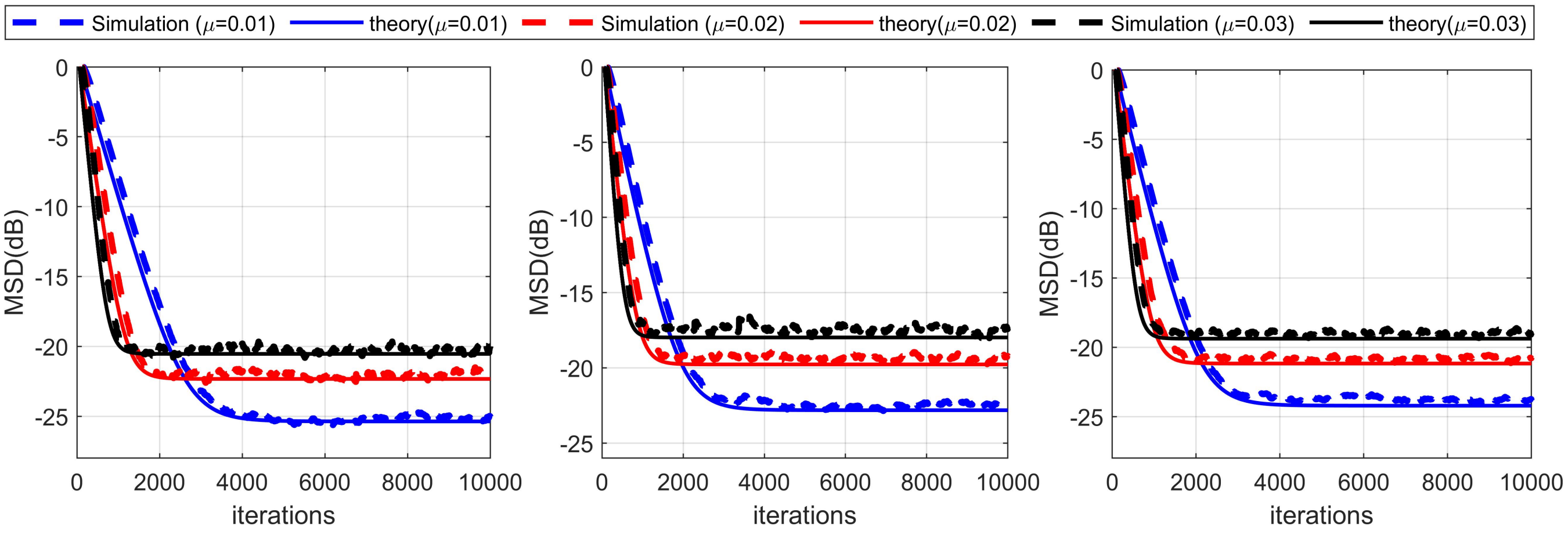

(a) (b) (c)

Fig. 6 Theoretical and simulated transient network MSD curves under different step sizes. The number of subband is fixed to $N_D = 8$.

(a) Gaussian input signal. (b) AR (1) input signal. (c) AR (2) input signal.

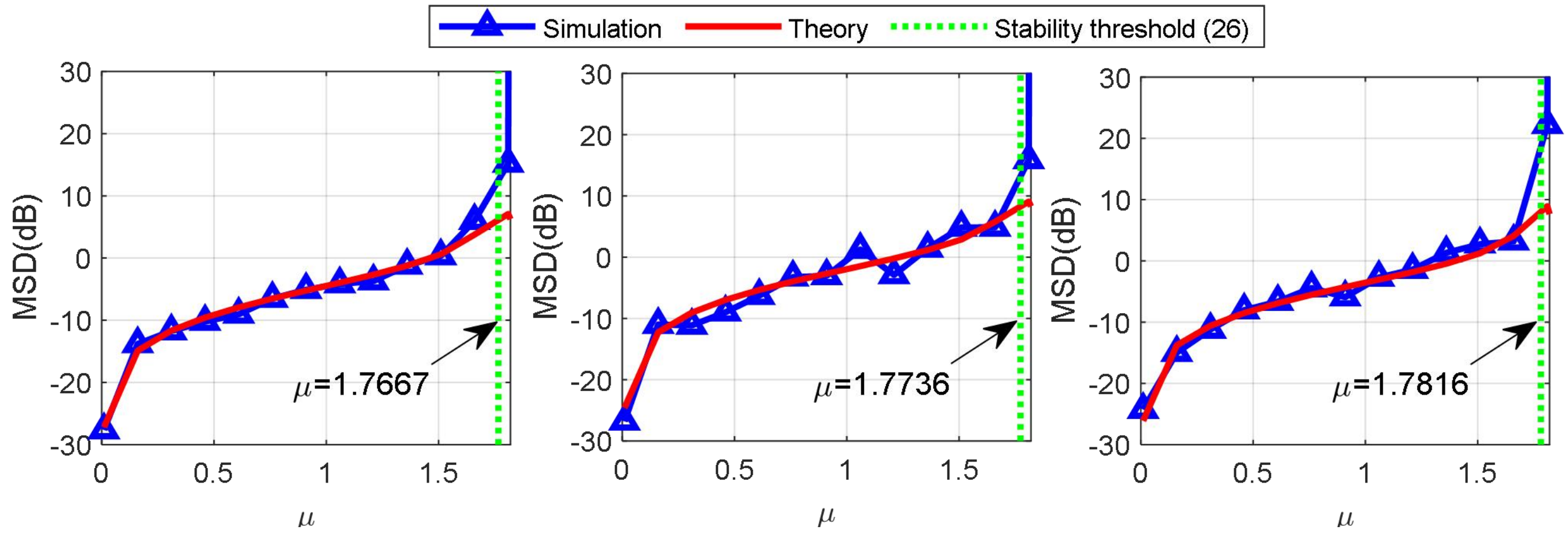

Fig. 7 Theoretical and simulated steady-state network MSD versus step size and the stability threshold given in (26).

(a) Gaussian input signal. (b) AR (1) input signal. (c) AR (2) input signal.

TABLE II

Parameters settings of all algorithms in performance comparison

| | Gaussian input signal | | | | | AR (1) input signal | | | | | AR (2) input signal | | | | |
|---|---|---|---|---|---|---|---|---|---|---|---|---|---|---|---|
| Algorithms | $\mu$ | $\eta$ | $N_D / P$ | $\sigma$ | $p_r$ | $\mu$ | $\eta$ | $N_D / P$ | $\sigma$ | $p_r$ | $\mu$ | $\eta$ | $N_D / P$ | $\sigma$ | $p_r$ |
| MD-APA | 0.008 | 0.01 | 2 | — | 0.01 | 0.008 | 0.01 | 2 | — | 0.001 | 0.008 | 0.01 | 2 | — | 0.01 |
| MD-APM | 0.009 | 0.01 | 2 | — | 0.01 | 0.009 | 0.01 | 2 | — | 0.001 | 0.0065 | 0.01 | 2 | — | 0.01 |
| MD-APMCC | 0.008 | 0.01 | 2 | 4 | 0.01 | 0008 | 0.01 | 2 | 4 | 0.001 | 0.008 | 0.01 | 2 | 4 | 0.01 |
| MD-NMSAF | 0.017 | 0.01 | 4 | — | 0.01 | 0.015 | 0.01 | 4 | — | 0.001 | 0.018 | 0.01 | 4 | — | 0.01 |

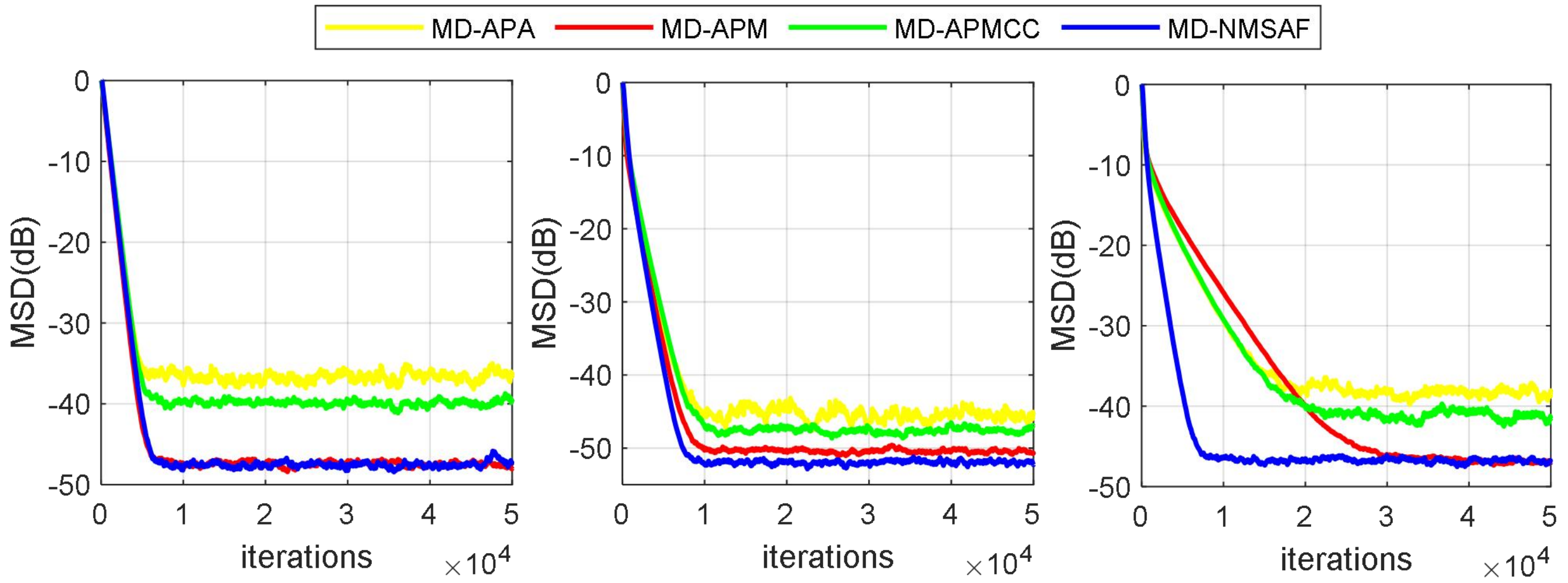

(a)

(b)

(c)

Fig.8 Convergence performances of the multitask diffusion algorithms. (a) Gaussian input signal. (b) AR (1) input signal. (c) AR (2) input signal.

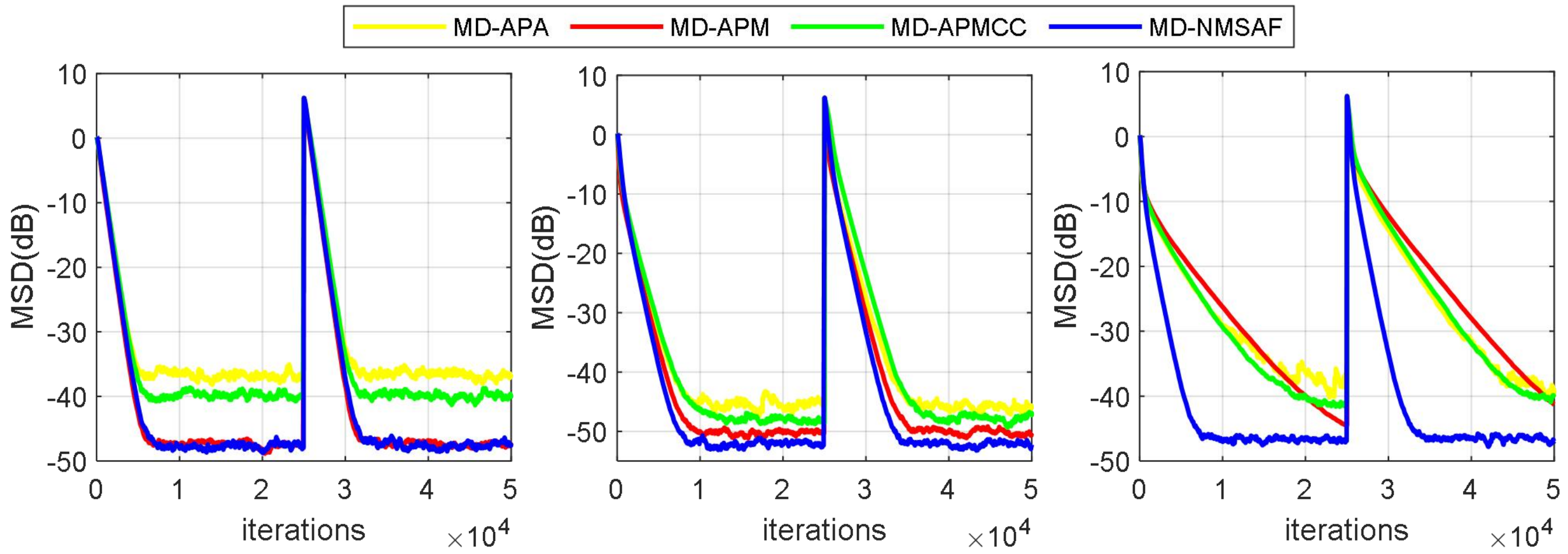

(a)

(b)

(c)

Fig.9 Tracking performances of the multitask diffusion algorithms. (a) Gaussian input signal. (b) AR (1) input signal. (c) AR (2) input signal.

## 5. Conclusions

In this paper, a novel robust multitask adaptive algorithm called MD-NMSAF is proposed, which inherits the strong decorrelation ability of the SAF for correlated inputs and the robustness of the M-estimate function. Compared with the AP based multitask adaptive algorithms, the MD-NMSAF algorithm has greatly improved the convergence speed and steady-state accuracy. More importantly, the computational complexity of the MD-NMSAF algorithm is much less than those of multitask AP algorithms. In addition, the stability condition, analytical expressions of the transient and steady-state network MSD of the MD-NMSAF algorithm in impulsive noise environment are provided and verified by computer simulations under different input signals, number of subbands and step sizes. Finally, the performance comparison between the MD-NMSAF algorithm and some other multitask algorithms fully confirms the performance advantages of the proposed algorithm.

## 6. Acknowledgment

This work was partially supported by National Natural Science Foundation of China (grant: 62171388, 61871461, 61571374), Fundamental Research Funds for the Central Universities (grant: 2682021ZTPY091), and the funding of Chengdu Guojia Electrical Engineering Co.,Ltd (grant: NEEC-2019-A02).